\documentclass[epj]{webofc}
\usepackage[varg]{txfonts}   % Web of Conferences font
\usepackage[abs]{overpic}
\woctitle{MESON2014 - 
the 13$^\textrm{th}$ International Workshop on Meson Production, Properties and Interaction}
\begin{document}
\selectlanguage{english}
\title{Diffractive production of mesons}

% insert email only for speaker/presenter
\author{Rainer Schicker\inst{1}\fnsep\thanks{\email{schicker@physi.uni-heidelberg.de}} 
% comment out the next line if not needed
%       \\for the XXXXX Collaboration
}

\institute{Phys. Inst., University Heidelberg }

\abstract{%
The interest in the study of diffractive meson production is discussed.
The description of diffraction within Regge phenomenology is presented,
and the QCD-based understanding of diffractive processes is given. 
Central production is reviewed, and the corresponding main results from 
the COMPASS experiment and from the experiments at the ISR, RHIC, TEVATRON and 
LHC collider are summarised. 
}
\maketitle
\section{Introduction}
\label{intro}

The production of mesons is interesting for a variety of reasons. First, meson production at 
a hard scale has a description within perturbative Quantum-Chromo-Dynamics (QCD) with quarks and 
gluons being the relevant degrees of freedom. For meson production at a soft scale, hadronic 
degrees of freedom need to be invoked. Here, Regge phenomenology has 
been the traditional framework with Regge trajectories representing the hadronic states.
Second, diffractive meson production is of particular interest as a tool to investigate 
the role of multi-gluon colour-singlet exchanges in strong interactions.
The dynamics of these multi-gluon singlet exchanges is so far only
poorly understood within QCD. Third, the fusion of multi-gluon objects 
is a gluon-rich environment, with very much suppressed quark degrees-of-freedom.
It is therefore anticipated that the hadronisation of these multi-gluon objects 
populates preferentially gluon-rich hadronic bound states, glueballs and hybrids.

The above issues can be addressed by studying central meson production at hadron colliders. 
Experimentally, central production is characterised by the very forward scattered beam 
particle, or remnants thereof, and by gaps in rapidity between the beam particle and the 
mid-rapidity system. A gap is characterised by the property that no particles are produced 
within this rapidity interval. Centrally produced systems are hence often characterised as 
double-gap events due to the presence of two gaps, one on either side of mid-rapidity.
The study of centrally produced systems at different centre-of-mass energies reveals
the contribution of colour-singlet quark and colour-singlet gluon exchanges.
These two different exchanges have a different energy dependence, as will be
discussed below. 

The study of glueballs and hybrids is based on the analysis of centrally produced 
resonances decaying into pion or kaon pairs. A Partial Wave Analysis (PWA) of the decay 
distributions of these pion and kaon pairs reveals the quantum numbers $J^{PC}$ of 
the decaying state. States with quantum 
numbers \mbox{$J^{PC}$ = $0^{--},0^{+-},1^{-+},2^{+-}$} cannot be $q\bar{q}$-mesons, hence
are of high interest for hadron spectroscopy. Such exotic states can be of 
tetraquark nature ($q\bar{q} + \bar{q}q$), or gluonic hybrids ($q\bar{q}$ + gluon).
In addition, such a decomposition into states of known quantum numbers will bring
new information on the existence of super-numerous states in the scalar sector \cite{Ochs}. 

\section{Diffraction}
\label{sec-1}

The phenomenon of diffraction has been known in optics for a long time due to the 
systematic studies by Joseph von Fraunhofer (1787-1826) and Augustin-Jean Fresnel (1788-1827).
The reconstruction of the diffraction figure based on Maxwell's equation was achieved by Gustav 
Kirchhoff (1824-1887). A plane wave of light passing through a circular hole or a slit
will form a diffraction pattern on a screen located behind the hole or slit. 
This pattern consists of a series of diffraction maxima and minima, which can be understood
to arise from Huygens principle describing the wave propagation. 
The first use of the term diffraction in nuclear high-energy physics was introduced 
by Landau and Pomeranchuk (1953).  Good and Walker formulated an anticipation in 1960: 

``A phenomenon is predicted in which a high-energy beam particle undergoing diffraction scattering
from a nucleus will acquire components corresponding to various products of the virtual dissociations
of the incident particle. These diffraction-produced systems would have a characteristic extremely
narrow distribution in transverse momentum, and would have the same quantum numbers as the initial
particle.''

Diffractive reactions in hadronic collisions can be characterised by two equivalent formulations:

\begin{itemize}  

\item a reaction in which no quantum numbers are exchanged between the colliding particles
is, at high energy, a diffractive reaction; 
 
\item a diffractive reaction is characterised by a large, non-exponentially suppressed, rapidity
gap in the final state.

\end{itemize}  

In the Regge framework, hadronic interactions are described by an exchange of objects,
the Reggeons, which are characterised by their so-called trajectory. Hadronic interactions
at high energies are dominated by the Pomeron trajectory. The different contributions
of Reggeon and Pomeron exchanges are clearly visible in the energy dependence of
hadronic cross sections \cite{DL}. A variety of event topologies can result from these 
exchanges.

\begin{minipage}[t]{.98\textwidth}
\begin{overpic}[width=.26\textwidth]{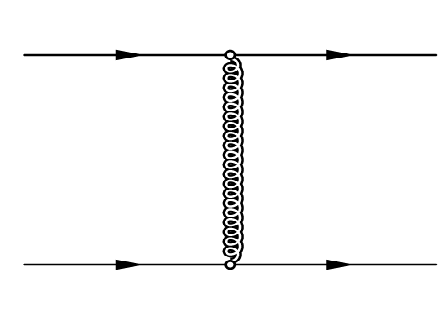}
\put(30.,1.){\footnotesize elast. scattering}
\end{overpic}
\hspace{1.cm}
\begin{overpic}[width=.28\textwidth]{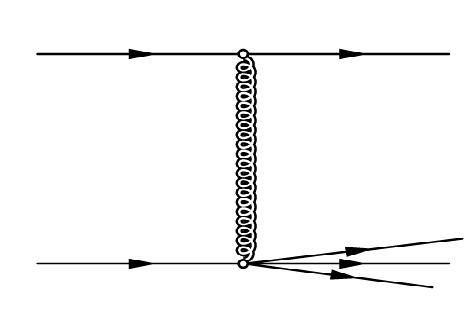}
\put(30.,1.3){\footnotesize single diff. diss.}
\end{overpic}
\hspace{1.cm}
\begin{overpic}[width=.26\textwidth]{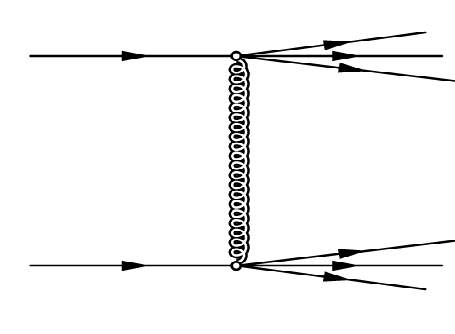}
\put(30.,1.){\footnotesize double diff. diss.}
\end{overpic}
\end{minipage}

\begin{minipage}[t]{.98\textwidth}
\begin{overpic}[width=.26\textwidth]{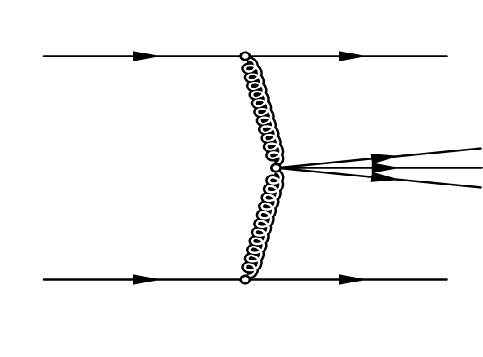}
\put(30.,1.){\footnotesize central prod.}
\end{overpic}
\hspace{1.1cm}
\begin{overpic}[width=.26\textwidth]{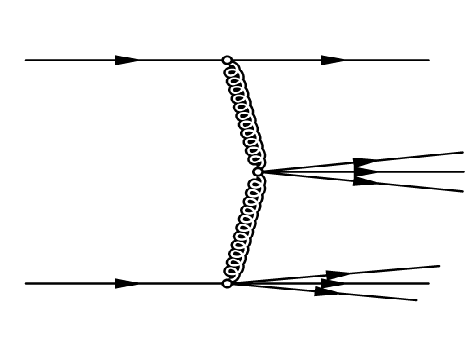}
\put(10.,1.){\footnotesize central prod./single diss.}
\end{overpic}
\hspace{1.cm}
\begin{overpic}[width=.26\textwidth]{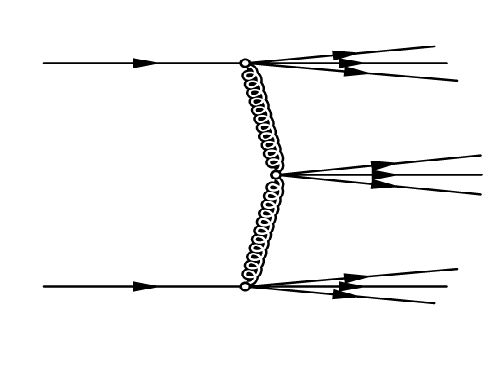}
\put(10.,1.){\footnotesize central prod./double diss.}
\end{overpic}
\end{minipage}
\begin{figure}[ht]
\vspace{-.5cm}
\caption{Diffractive event topologies.}
\label{fig:1}       % Give a unique label
\end{figure}

Figure \ref{fig:1} displays event topologies with rapidity gaps.
In addition to the Reggeon/Pomeron exchanges discussed above, photon and 
W$^{\pm}$\,-\,exchanges contribute to these topologies. Due to the large photon flux
present in heavy-ion beams, in particular photon contributions need to be taken
into account with heavy-ion beams. Reggeon/Pomeron and photon exchanges contribute, 
however, differently in pp, pA and AA-collisions. A systematic study of meson 
production in these systems is therefore mandatory for a comprehensive 
understanding of the underlying production mechanisms.  

\section{Central production}
\label{sec-2}

The topologies in the lower part of Figure \ref{fig:1} are of particular
interest for the study of diffractive meson production. The hadronisation  of two 
Reggeon/Pomeron exchanges can take place with and without dissociation of the beam 
particle as shown in the three bottom diagrams. In the case of approximate
energy sharing between the two exchanges, the central system will be formed
close to mid-rapidity.  

In the Regge regime ($\sqrt{s} \rightarrow \infty, \sqrt{\,|t|} \leq$  1 GeV, t = square 
of four-momentum transfer), the modeling
of such reactions is based on exchanges of the Pomeron and Reggeons $f_2$, $a_2$,
$\omega,\rho$. A fundamental question is the nature of the Pomeron.
Recent theoretical work resulted in the formulation of a model combining general rules
of Quantum Field Theory (QFT) with Regge theory.  This model aims at giving simple
rules for calculating exchange amplitudes compatible with QFT. First results within this 
model include new insights into the meaning of Vector-Dominance relations, relations
between particle-particle-particle and Reggeon-particle-particle vertices,
and the appearance of the Pomeron as an effective rank-two tensor exchange \cite{Nachtmann}.
A report on exclusive central diffractive production of scalar, pseudoscalar and 
vector mesons within this model is given at this conference \cite{Lebie}.   
An investigation of exclusive photoproduction of J/$\Psi$ and $\Psi^{'}$ 
is presented at this conference \cite{Cisek}.

Early searches for double-gap-topology events  were carried out at the Serpukov accelerator 
at \mbox{$\sqrt{s}$ = 11.5 GeV.} All the events found were consistent with single-diffractive 
dissociation, and hence only an upper limit on the central exclusive cross-section could be 
derived \cite{Serpukov}. 

\subsection{Central production at the ISR}

The first observation of double-gap events was achieved by the ARCGM Collaboration
at the Intersecting Storage Ring (ISR) at CERN \cite{ARCGM}. This experiment measured the 
forward protons and had full angular coverage with scintillation counters.  Measurements of 
centrally produced $\pi^{+}\pi^{-}$-pairs were accomplished at the ISR by the CCHK 
Collaboration at $\sqrt{s}$ = 31 GeV \cite{DellaNegra}, and later by the  Axial Field 
Spectrometer Collaboration  with much improved statistics at $\sqrt{s}$ = 63 GeV \cite{Akesson}.  

\vspace{.2cm}
\begin{minipage}[h]{.62\textwidth}
\begin{overpic}[width=.94\textwidth]{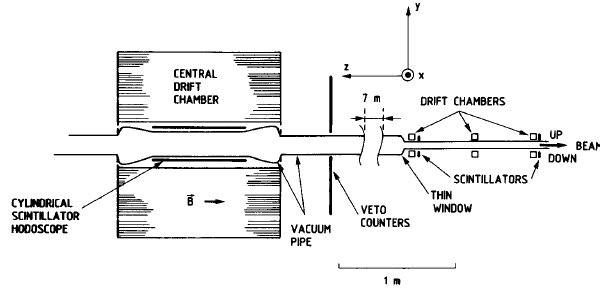}
\put(36.,106.){\footnotesize Axial Field Spectrometer}
\put(1.,8.){\footnotesize Apparatus is left-right symmetric}
\put(1.,-2.){\footnotesize forward detectors shown only on right-hand side}
\end{overpic}
\hspace{.8cm}
\vspace{.8cm}
\end{minipage}
\begin{minipage}[h]{.44\textwidth}
\vspace{-1.4cm}
\hspace{-.4cm}
\begin{overpic}[width=.78\textwidth]{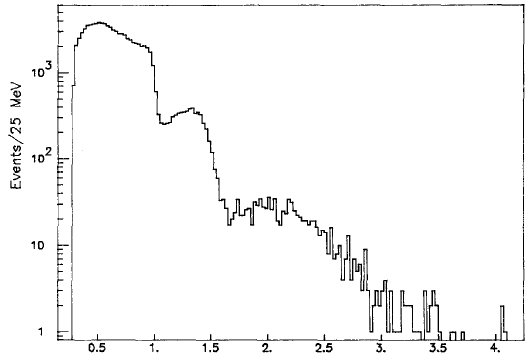}
\put(50.,98.){\footnotesize exclusive $\pi^{+}\pi^{-}$}
\put(100.,-8.){\footnotesize M($\pi\pi$) GeV}
\end{overpic}
\end{minipage}
\begin{figure}[ht]
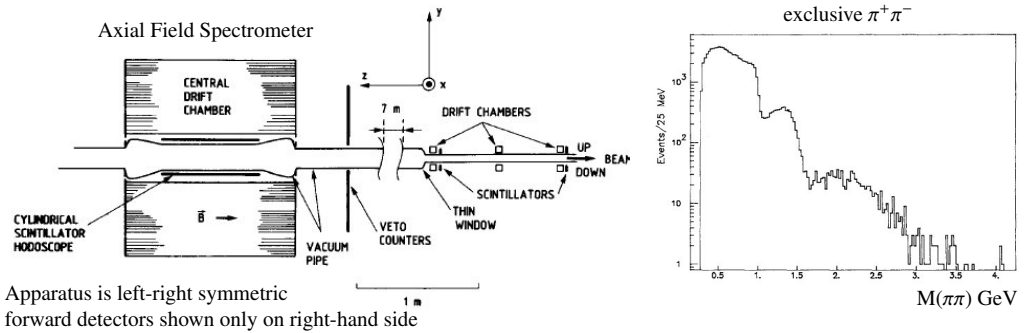

\vspace{-1.2cm}
\caption{The Axial Field Spectrometer at the ISR (taken from Ref. \cite{Akesson}).}
\label{fig:2}       % Give a unique label
\end{figure}
   
Figure \ref{fig:2} shows the Axial Field Spectrometer setup on the left. The central 
detector consists of a drift chamber and of cylindrical scintillator hodoscopes.
The information from the Veto counters establishes the rapidity gap. In forward direction, 
drift chambers and scintillators register
the forward-scattered beam particles. On the right side of Figure \ref{fig:2}, 
the invariant-mass spectrum of pion pairs measured at mid-rapidity is shown.
Clear peaks associated to resonances are seen. 
Due to the quantum numbers $J^{PC}$ = 1$^{--}$, the $\rho$-resonance can be produced
when a Reggeon exchange is involved, but not in pure Pomeron-Pomeron exchange.
The absence of the $\rho$-signal in this spectrum is therefore evidence that 
Reggeon exchanges contribute very little at the energy of $\sqrt{s}$ = 63 GeV.

\subsection{Central production at the COMPASS experiment}

COMPASS is a fixed-target experiment at the CERN SPS accelerator. The experimental setup
consists of a two-stage magnetic spectrometer, of tracking detectors for momentum reconstruction 
and of calorimeters and RICH detectors for particle identification. The data taken by COMPASS 
in the years 2008 and 2009 allow the analysis of central meson production. A comparison of the
COMPASS data at $\sqrt{s}$ = 18.9 GeV to the data at the ISR experiments clearly shows the 
gradual disappearance of the centrally produced $\rho$-signal in the energy range from 
$\sqrt{s}$ = 12.7 GeV to $\sqrt{s}$ = 63 GeV, and hence is evidence for the interplay between 
Reggeon and Pomeron exchanges.  The present status of the COMPASS partial wave
analysis of centrally produced pion and kaon pairs is presented at this conference \cite{COMPASS}.

\subsection{Central production at STAR}

The STAR experiment at the Relativistic Heavy Ion Collider (RHIC) at Brookhaven consists of a Time 
Projection Chamber (TPC), a silicon vertex tracker, electromagnetic calorimetry and scintillator 
paddles for time-of-flight measurements. The central TPC with a pseudorapidity coverage 
$ -1 < \eta < 1$ allows the measurement of the central system, while the rapidity gap is defined 
by the information of the Forward TPC (FTPC, $ 2.5 < |\eta| < 4.2$) and the beam-beam counters 
(BBC, $ 3.8 < |\eta| < 5.2$). In addition, Roman Pots installed at a distance of 55.5\,m and 58.5\,m 
from the interaction region allow the measurement of the forward-scattered protons.  In Phase I of 
STAR running at \mbox{$\sqrt{s}$ = 200 GeV}, the acceptance of the Roman Pots allowed the measurement 
of forward protons in the range \mbox{$ 0.003 < |t| < 0.035$ GeV$^2$}. 
In accepted events, the kinematics of all the final-state particles is known.
This information of the final state can be used to efficiently remove non-exclusive background 
by imposing a condition on the transverse momentum balance $p_{T}^{miss} < 0.02$ GeV.  
The STAR analysis of central production from data taken in 2009 resulted in  380 clean events,
yielding differential cross sections in the invariant mass and rapidity of the centrally 
produced pion pairs \cite{STAR}.

\subsection{Central production at the TEVATRON}

The Tevatron collider at Fermilab provided proton-antiproton collisions in Run I (1992-1996) at
energies $\sqrt{s}$ = 546, 630 and 1800 GeV. In this Run, the Collider Detector (CDF) had Roman Pots 
installed, and was thereby able to record  elastically and inelastically scattered protons and antiprotons. 
From these data, the elastic and the total cross section could be derived. Unfortunately, these Roman 
Pots were removed before central production could be measured in coincidence with the forward beam 
particles. In Run II (2004-2011), the CDF Collaboration installed a new Roman Pot for measuring 
antiprotons, and scintillation counters were installed as beam shower counters in the range 
$5.4 < |\eta| < 7.4$. In conjunction with miniplug calorimeters, a rapidity gap $\Delta\eta > 3.8$ 
could be required on either side of mid-rapidity. The data were analysed  by imposing this gap 
condition in addition to the Roman Pot information. From this analysis, results were derived for 
diffractive dijet and diffractive W-boson production, exclusive $\gamma\gamma$ and exclusive 
$\chi_{c}$-production\cite {Tevatron1,Tevatron2}.

The CDF Collaboration implemented special triggers during the Tevatron energy scan running at
$\sqrt{s}$ = 300 and 900 GeV in September of 2011. These triggers included  zero-bias and minimum-bias 
conditions, a jet trigger, a lepton trigger as well as a Gap-X-Gap trigger for the double 
\mbox{gap topology \cite{CDF}}. The present status of the ongoing analysis is presented at this
conference \cite{MariaZ}.

\subsection{Central production at the LHC}

\subsubsection{Central production at ALICE}

The ALICE experiment at the Large Hadron Collider (LHC) at CERN consists
of a central barrel, a muon spectrometer and of additional detectors for 
trigger and event classification purposes. The low transverse-momentum
threshold of the central barrel gives ALICE a unique opportunity to study
the low-mass sector of central exclusive production at the LHC.

The ALICE Collaboration has taken data in proton-proton, proton-lead as well as 
lead-lead collisions in Run I of the LHC. 
In the years 2010-2011, ALICE recorded zero-bias and minimum-bias data in 
pp-collisions at a centre-of-mass energy of $\sqrt s$ = 7 TeV. 
Events with double-gap topology are contained in the minimum-bias trigger, 
hence central diffractive events were analysed from the minimum bias data sample. 
The multiplicity distributions of the double and  no-gap events 
clearly show different behaviour, as discussed in Ref. \cite{Schicker}.

In lead-lead collisions, exclusive  photoproduction of vector mesons is of 
particular interest. The vector meson production cross-section depends 
on the nuclear gluon distribution, hence allows the study of nuclear 
gluon-shadowing effects at values of Bjorken-x $\sim 10^{-3}$ and 
$\sim 10^{-2}$ for data taken in the ALICE central barrel and 
the muon spectrometer, respectively.
A report on the ALICE results in coherent photo-production
of $\rho^{0}$-mesons in ultra-peripheral lead-lead collisions
is given at this conference \cite{Mayer}.

\subsubsection{Central production at LHCb}

The LHCb experiment consists of a single-arm forward spectrometer covering the pseudorapidity
range $2 < \eta < 5$, of a tracking system surrounding the interaction region (VELO),
and three stations of silicon-strip detectors and straw drift-tubes placed downstream
of the magnet. Charged hadrons are identified using two ring-imaging Cherenkov detectors.
Photon, electron and hadron candidates are identified by a calorimeter system consisting
of scintillating-pad (SPD) and pre-shower detectors, an electromagnetic and hadronic calorimeter.
Rapidity gaps can be established by the VELO detector. A requirement of exactly 
two muon tracks within the spectrometer acceptance hence allows the study of exclusive quarkonia
production\cite{LHCb}.  

\begin{figure}[ht]
\centering
\includegraphics[width=6.9cm,clip]{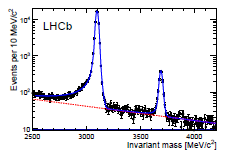}
\caption{Mass spectrum of exclusively produced dimuons (taken from Ref. \cite{LHCb}).}
\label{fig:3}       % Give a unique label
\end{figure}

Figure \ref{fig:3} shows the dimuon invariant-mass spectrum measured in the acceptance 
of LHCb. Clearly visible are the J/$\Psi$ and the $\Psi$(2S) signals, as well as a continuum arising 
from $\gamma\gamma \rightarrow \mu^{+}\mu^{-}$. From these data, cross sections for
exclusive production of J/$\Psi$ and $\Psi$(2S) can be derived \cite{LHCb}.
 
\subsubsection{Exclusive  production at CMS}

The Compact Muon Solenoid (CMS) experiment consists of a silicon tracker with pseudorapidity
coverage $|\eta| < 2.5$, of electromagnetic and hadronic calorimeters in the range $|\eta| < 3$
and of a forward hadronic calorimeter with coverage $ 2.9 < |\eta| < 5.2$.  
The CASTOR detector extends the CMS coverage on one side to the region $5.3 < |\eta| < 6.6$.
An overview of the CMS  results on exclusive production is given at this conference \cite{CMS}.

\section{Conclusion}

A wealth of data is available for central meson production at hadron colliders, from
the early low energies of the ISR to the presently highest energies of the LHC. A partial 
wave analysis of the decay distributions of pion and kaon pairs reveals the quantum numbers
of the centrally produced resonances. Such information allows the study of exotic
mesonic states, a topic of fundamental interest in meson spectroscopy.

\begin{acknowledgement}
This work is supported by the German Federal Ministry of Education and 
Research under promotional reference 05P12VHCA1 and by WP8 of the hadron 
physics programme of the 7th and 8th EU programme period.
\end{acknowledgement}

\end{document}